\documentclass[preprint,preprintnumbers,amsmath,amssymb,
nofootinbib,eqsecnum,12pt]{revtex4}

\usepackage{amsmath}
\usepackage{graphicx}

\newlength{\dummysp}
\newcommand{\beq}{\begin{eqnarray}}
\newcommand{\eeq}{\end{eqnarray}}

\newcommand{\s}{{\sigma}}

\newcommand{\gappeq}{\mathrel{\rlap {\raise.5ex\hbox{$>$}}
{\lower.5ex\hbox{$\sim$}}}}
\newcommand{\lappeq}{\mathrel{\rlap{\raise.5ex\hbox{$<$}}
{\lower.5ex\hbox{$\sim$}}}}

\newcommand{\ben}{\begin{enumerate}}
\newcommand{\een}{\end{enumerate}}

\newcommand{\bit}{\begin{itemize}}
\newcommand{\eit}{\end{itemize}}

\newcommand{\susy}{SUSY}

\newcommand{\Ncal}{{\cal N}}

\newcommand{\FKV}{Fleming et al.~\cite{Fleming:2000fa}}
\newcommand{\DMR}{DESY-M\"unster-Roma}

\newcommand{\mres}{m_{\text{res}}}

\def\[{\left [}
\def\]{\right ]}
\def\({\left (}
\def\){\right )}

\begin{document}

\title{Gluinos condensing at the CCNI:  4096 CPUs weigh in\footnote{Talk given by
Joel Giedt at ``Continuous Advances in QCD 2008,'' Fine Theoretical Physics
Institute, University of Minnesota, Minneapolis, MN, May 15-18, 2008.}}

\author{Joel Giedt}
\email{giedtj@rpi.edu}
\affiliation{Department of Physics, Applied Physics and Astronomy, \\
Rensselaer Polytechnic Institute, \\ 110 8th Street, Troy NY 12065 USA}

\author{Richard Brower}
\email{brower@bu.edu}
\affiliation{Physics Department, Boston University \\
590 Commonwealth Avenue, Boston MA 02215}

\author{Simon Catterall}
\email{smc@syracuse.edu}
\affiliation{Department of Physics, Syracuse University, Syracuse, NY 13244 USA}

\author{George T. Fleming}
\email{George.Fleming@yale.edu}
\affiliation{Department of Physics, Sloane Laboratory, 
Yale University, New Haven, Connecticut 06520, USA}

\author{Pavlos Vranas}
\email{vranas@llnl.gov}
\affiliation{Physical Sciences Directorate,
Lawrence Livermore National Laboratory, 7000 East Ave., Livermore, CA 94550}

\date{July 12, 2008}

\begin{abstract}
We report preliminary results of lattice super-Yang-Mills computations 
using domain wall fermions, performed at an actual
rate of 1000 Gflop/s,
over the course of six months, using two BlueGene/L racks at
Rensselaer's CCNI supercomputing center.
This has allowed us to compute the
gluino condensate and string tension over a wide
range of lattice parameters, setting the stage
for continuum, chiral extrapolations.
\end{abstract}

\maketitle

%
In this talk, I present preliminary results obtained in
collaboration with coauthors of a forthcoming paper \cite{Gie08a}.
We have
used domain wall fermions (DWF) \cite{Kaplan:1992bt,Shamir:1993zy}
to study nonperturbative aspects of
pure ${\cal N}=1$ super-Yang-Mills (SYM) \cite{Ferrara:1974pu}.
The strong dynamics of supersymmetric gauge theories underlie most models
of spontaneous supersymmetry (SUSY) breaking, and the development
of a first-principles tool is needed.
Though at finite lattice spacing \susy\ is violated,
it is automatically recovered \cite{Curci:1986sm} in the
continuum limit with a massless gluino.\footnote{Here we use
the terms ``gluon'' and ``gluino'' by way of analogy.  It should
be kept in mind that the strongly coupled gauge theory would be
an extension to the gauge group of the Standard Model.}
When the DWF formulation is
employed, this ``chiral limit'' is achieved without the
need for a computationally expensive, 
nonperturbatively determined fine-tuning of the bare
gluino mass \cite{Neuberger:1997bg,Kaplan:1999jn};
in the limit of infinite domain wall separation
$L_s \to \infty$, DWF realize
the lattice chiral symmetry \cite{Luscher:1998pqa} associated with
Ginsparg-Wilson fermions \cite{Ginsparg:1981bj}, which protects
against additive mass renormalization.  
These nice features of the DWF
approach are to be contrasted with the Wilson fermion formulation,
which was pursued for several years by the \DMR\ collaboration \cite{Campos:1999du,
Montvay:2001aj,Peetz:2002sr,Farchioni:2001wx,
Farchioni:2004ej,Farchioni:2004fy,Pee03}.

Lattice studies can provide details that other approaches cannot,
such as ``snapshots'' of the gauge field configurations that
are dominating the gluino condensate.  For instance, in the
work of \FKV, the only DWF simulation
of SYM to date, it was suggested that spikes in the 
gluino condensate may correspond to configurations 
with fractional topological charge,
as would be expected from Ref.~\cite{Davies:1999uw}.
Further pursuit of this conjecture, consisting of lattice studies of
monopoles and topological charge, is on our agenda.
Also,  we will
compute the low-lying spectrum of composite states, consisting
of strongly bound gluons and gluinos.  Apart from the inconclusive
results of \DMR, this aspect of SYM is completely unknown from
continuum methods, and ideally suited to the lattice approach.

At early stages in such studies, 
understanding numerical behavior
of important quantities such as the gluino condensate will
teach us a lot about the lattice formulation that we currently
do not know.  For instance, it is important to set ``benchmarks''
regarding compute time, 
and lattice artifacts such as discretization
and finite size effects.  The first place that we will
examine this is in the gluino condensate, which is believed
to be known exactly by continuum methods \cite{Affleck:1983mk,
Affleck:1984xz,Novikov:1985ic,Davies:1999uw,Cachazo:2002ry}, and
is therefore ideal for calibrating the lattice methods.  Understanding of the
lattice theory and simulation performance is already emerging from
our preliminary results, as we now briefly discuss.

Domain wall fermion simulations require
world-class computing resources, such as are
are available to Giedt at Rensselaer; namely, the
Computational Center for Nanotechnology Innovations (CCNI),
one of the world's most powerful university-based 
supercomputing centers, and a top 25 supercomputing
center of any kind in the world.  We are presently the
third heaviest user of this facility, and have been generating
lattice configurations and measurements continuously at a sustained actual
rate of 1000 Gflop/s since the end of January 2008.  For comparison
the \DMR\ collaboration performed their computations at
a rate of 10 Gflop/s for a cumulative time of one year.
Thus our study represents a hundred-fold improvement over
what has been done previously, just in terms of raw computation
power.

\begin{figure}
\begin{center}
\includegraphics[width=3in,height=5in,angle=90]{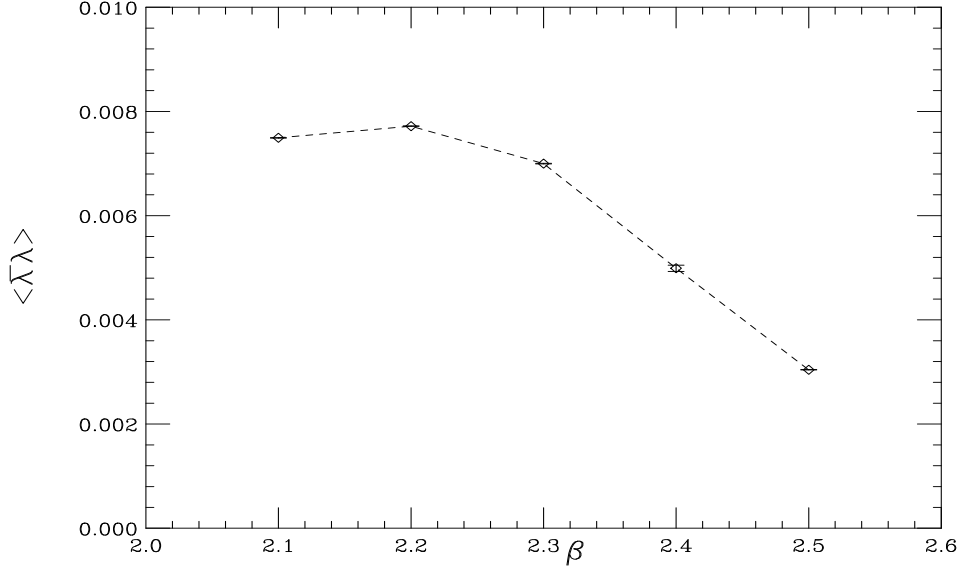}
\caption{
The gluino condensate versus~$\beta$ for a
$16^3 \times 32$ lattice with
domain wall separation $L_s=16$ (dashed line drawn to guide the eye).
It can be estimated from the figure that for the $16^3 \times 32$ lattice studied here,
the system will deconfine at $\beta \sim 2.7$, as a result of finite
size effects.}
\label{cf1}
\end{center}
\end{figure}

As an example of our results, we have obtained the bare gluino condensate
from dynamical domain wall fermion simulations for a
variety of bare gauge couplings $g$, parameterized in terms
of $\beta=4/g^2$, as is conventional in SU(2) lattice gauge theory.
The results for a $16^3 \times 32$ lattice
(i.e., the number of sites in spatial and temporal directions)
with domain wall separation $L_s=16$ sites are displayed in
Fig.~\ref{cf1}.  We note that such data for the condensate
versus $\beta$ has never been obtained before; it is important
because the continuum limit corresponds to $\beta \to \infty$ (with
the physical size of the lattice held fixed).
A fit to the data obviously yields a vanishing condensate at 
a finite gauge coupling $\beta \sim 2.7$.  This just reflects
the fact that as $\beta$ increases the lattice spacing shrinks,
and thus so does the physical size of the lattice in its
entirety.  In a small enough ``box'' confinement will
disappear and the condensate ``melts.''  Thus we already
gain an important benchmark:  to go much beyond $\beta=2.5$
will require larger lattices, and in fact one should
carefully measure systematic errors due to
finite size effects at $\beta\approx 2.5$.  This is
consistent with what is already known from the so-called
``quenched'' theory, which has no gluinos.

\begin{figure}
\begin{center}
\includegraphics[width=3in,height=5in,angle=90]{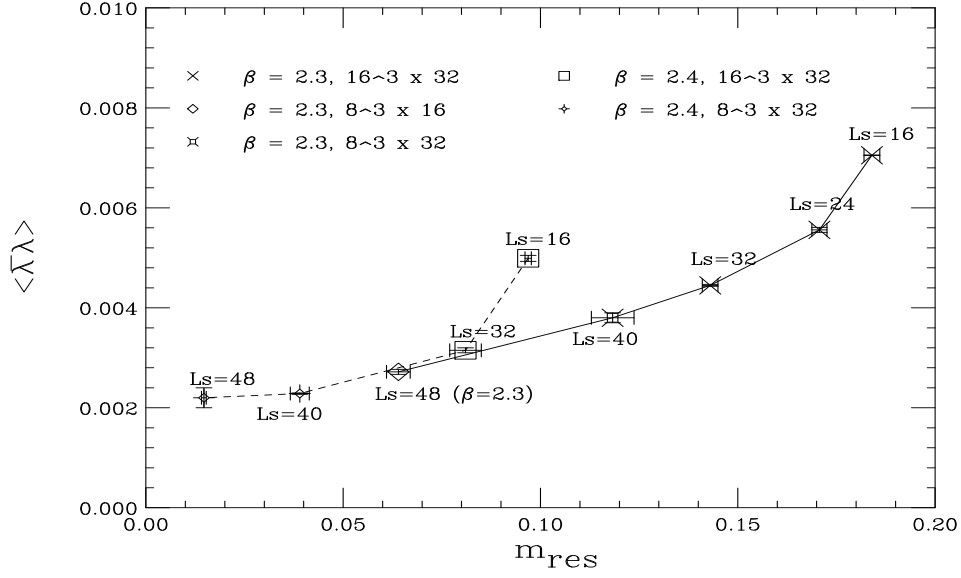}
\caption{
Our simulation results for the gluino condensate 
versus~$\mres$, where the latter is
a measure of explicit chiral symmetry breaking due to finite
domain wall separation $L_s$.  The solid line corresponds to $\beta=2.3$
whereas the dashed line is for $\beta=2.4$.}
\label{cf2}
\end{center}
\end{figure}

Fig.~\ref{cf2} shows the gluino condensate for decreasing values
of the residual mass $\mres$, which is a measure of
explicit chiral symmetry breaking due to finite $L_s$ \cite{Blum:2000kn}.
As expected, larger $L_s$ values have the smallest $\mres$,
and a nonzero gluino condensate appears to
occur in the $\mres \to \infty$ limit.  Also as expected,
smaller values of $\mres$ occur for the weaker coupling $\beta=2.4$.

\begin{figure}
\begin{center}
\includegraphics[width=3in,height=5in,angle=90]{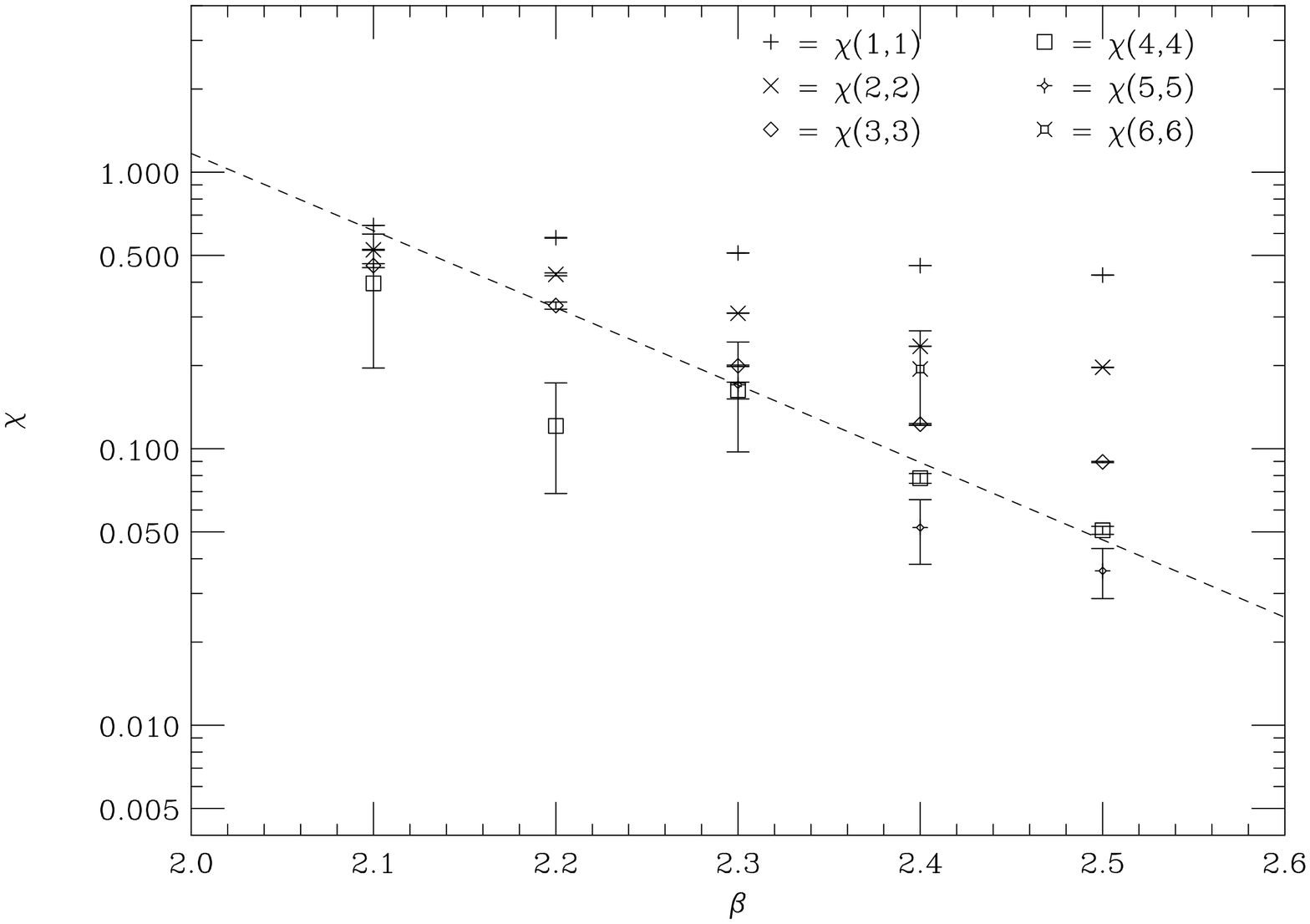}
\caption{Creutz ratios for the $16^3 \times 32$ lattice
with $L_s=16$.  The dashed line indicates the 2-loop \susy\ prediction for
the dependence $\chi \sim \s a^2$ where $\s$ is
the string tension and $a=a(\beta)$ is the lattice spacing. \label{cf16}}
\end{center}
\end{figure}

Finally, we have looked at Creutz ratios \cite{Creutz:1980wj},
\beq
\chi(R,R) = - \ln \frac{ W(R,R) W(R-1,R-1) }{W(R,R-1)^2} \sim  \s a^2 ,
\label{cre}
\eeq
where $W(R,R')$ is an $R \times R'$ Wilson loop,
in order to extract the string
tension $\s$ in lattice units, as well as to delineate the scaling
regime where the continuum limit may be extracted.  In the process
we obtain an estimate of the lattice spacing in units of the string tension.
Results for the $16^3 \times 32$, $L_s=16$ lattice are shown
in Fig.~\ref{cf16}.  Although the errors
are somewhat large, scaling is clearly setting in at around $\beta=2.3$.
To see this one notes that the larger Creutz ratios appear to
coalesce on an envelope, corresponding to the distance scale at
which an area law begins to take hold in the Wilson loops.

We emphasize that all of the results mentioned above
are ground-breaking as far as lattice SYM is concerned.
Several important goals related to the lattice SYM project
have already been achieved, laying the groundwork for the
more extensive studies that will follow:
\bit
\item
Developed parallel simulation code for SYM by modification
of the current version of the Columbia Physics System (CPS) QCD package.
\item
This extends DOE funded code (CPS, part of USQCD's SciDAC program)
to ``beyond the Standard Model'' physics, which
is a realization of one of the USQCD Collaboration
objectives \cite{scidac-bsm}.
\item
We have reproduced the results of \cite{Fleming:2000fa}
as a check on our code.
\item
Our software runs successfully on IBM's Bluegene
(BG) architecture, taking full advantage of BG specific communications
utilities.
\item
Developed Landau gauge-fixing and Fourier space propagator code
for adjoint fermion representations, essential for nonperturbative renormalization
of the condensate, in the RI/MOM scheme \cite{Martinelli:1994ty,Blum:2001sr}.
\item
Established timing and statistical uncertainty benchmarks.
For example, we have found that for small single-node 
volumes ($16^3 \times 32 \times L_s / 2048$ CPU's $ = 64 \times L_s$ sites
per CPU) the efficiency of the BG parallel code is 10 percent.
\eit

In the course of our studies, we intend to investigate other domain wall fermion
formulations (e.g., ``gap'' \cite{Vranas:2006zk} and ``Mobius'' 
\cite{Brower:2004xi}) as ways to approach
the chiral limit more quickly.  Also, we will implement recent
optimizations of fermion matrix inverters (to increase
efficiency) and improved  actions (to reduce systematic
errors).

\section*{Acknowledgements}
%
We benefited from a copy of the code that was used
in \cite{Fleming:2000fa}, and employed it as the basis for our
modifications to the current version of the Columbia Physics
System.  At various points JG benefited from technical
assistance provided by Chulwoo Jung (Brookhaven National Lab) and
Adam Todorski (SCOREC and CCNI at Rensselaer).  
The computational efforts on this project mainly
utilized the Computational Center for Nanotechnology Innovations (CCNI),
and JG expresses his appreciation for continuous access to that facility.
The project also, at times, utilized
the SUR BlueGene/L at Rensselaer, which is supported 
by NSF grant 0420703 entitled ``MRI: Acquisition of Infrastructure for Research 
in Grid Computing and Multiscale Systems Computation''
and a gift by the IBM Corporation of a BlueGene/L computer.
JG acknowledges support from Rensselaer faculty development funds.

\end{document}